# Zone-resolved photoelectronic scoping of the local bonding and electronic dynamics at the graphite skin with and without atomic vacancy and the associated graphene edge states

Chang Q Sun,[1,2] Yanguang Nie,[1] Jisheng Pan,[3] Xi Zhang,[1] S. Z. Ma,[1] Yan Wang,[4] Weitao Zheng[5]

[1]School of Electrical and Electronic Engineering, Nanyang Technological University, Singapore 639798;

[2]Faculty of Materials, Photoelectronics and Physics, Xiangtan University, Changsha 400073, China;

[3]Institute of Materials Research and Engineering, A*Star, Singapore 117602; [4] School of Information and Electronic Engineering, Hunan University of Science and Technology, Xiangtan 411201, China;

[5]School of Materials Science, Jilin University, Changchun 130012, China.

*Ecqsun@ntu.edu.sg; wtzheng@jlu.edu.cn*

## Abstract

A zone-resolved photoelectron spectroscopy (ZPS) has enabled us to gain the local and quantitative information and hence confirm our theoretical expectations on the bonding and electronic dynamics at graphite skin with and without atomic vacancy defects. The ZPS study has revealed: i) the 1s energy level of an isolated carbon atom is located at 282.57 eV, which shifts by 1.32 eV deeper upon diamond bulk formation; ii) the graphite surface bonds contract by 18% with 165% gain in energy compared with the C-C bond in the bulk diamond; the surface C 1s energy shifts 2.08 eV deeper from the 1s level of an isolated carbon atom; and iii) the defect bonds are ~26% shorter and 215% stronger with the binding energy shift of ~2.85 eV. An additional polarization peak centered at 1.28 eV below the C 1s level presents when the vacancy is formed. Associated with the scanning tunneling microscopy/spectroscopy observations and density functional theory calculations, the ZPS measurements clarify, for the first time, that the graphitic Dirac-Fermi polarons at atomic vacancy or graphene zigzag edge arise from the polarization of the unpaired dangling-bond electrons by the undercoordination-induced local densification and quantum entrapment of the bonding electrons.

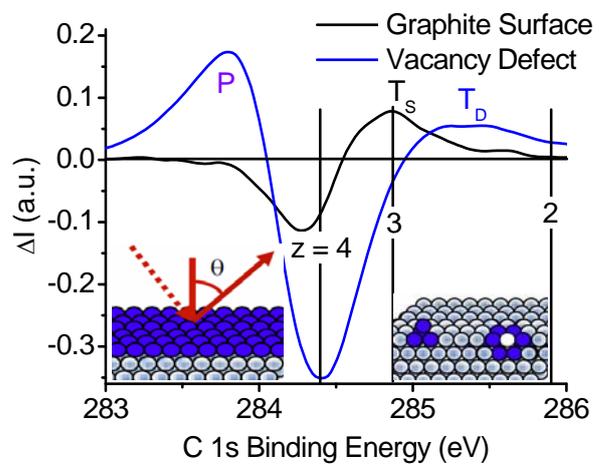

# I  Introduction

Bonds and electrons annexed the undercoordinated atoms dictate the unusual performance of materials at surface, defect, and at the nanoscale [1] in catalytic reactivity,[2, 3] crystal growth,[4] adsorption,[5] decomposition,[6] doping,[7] interface formation,[8, 9] wettability,[10] thermal stability,[11] etc. Although the chemistry and physics of materials associated with under-coordinated atoms have been extensively investigated for decades, the laws governing the performance of such local bonds and electrons remain as yet to be established.[12, 13, 14] Collection and purification of such local, atomistic zone selective, and quantitative information having been increasingly demanded but it remains as yet the "dead corner" of the community.

As a powerful detection tool, the scanning tunneling micro/spectroscopy (STM/S) maps local electrons in the open side of a *too-thin* subatomic layer of a surface with energies of a few eVs cross Fermi energy ($E_F$). Understanding the intriguing STM/S attributes of high protrusions and the additional resonant spectral peak nearby $E_F$ at graphite atomic vacancy [15] and graphene edge,[16] for instances, as well as the driving force for the unusual protrusions and the $E_F$ resonance remains challenge. One urgently needs to identify what the "root" of the STM/S attributes is and what the quantitative information could be about the length and strength of the annexed bonds and the energies of the associated electrons. On the other hand, a photoelectron spectroscopy (PES such as ultraviolet or x-ray as the sources called UPS and XPS, respectively) probes statistic and volumetric information of electrons with binding energy in the valence band and below within a *too-thick* layer of 10 nanometers or thicker.[13, 17, 18, 19] The interplay of STM/S and XPS could provide qualitative information of the surface and the bulk in a certain depth but with a challenge for localized, atomistic zone selective, quantitative information from a sheet of atomic-diameter-thick underneath what the STM/S can scope.

In order to solve this challenging issue, we have developed a special yet simple technique of zone-resolved photoelectron spectroscopy (ZPS), which has enabled us to overcome the aforementioned limitations. Applying the ZPS to the graphite surface with and without atomic vacancy, we have been able to derive important findings as elaborated in the following.

# II  Principles
## 2.1  The Hamiltonian determined core level shift

Firstly, according to the energy band theory,[20] the energy shift of a specific (the C 1s) core band from that of an isolated atom is proportional to the crystal potential energy at equilibrium or the cohesive energy per bond. Any perturbation to the crystal potential will shift the C 1s band away from the bulk value. The energy shift can be positive or negative depending on the perturbation resulting from bond relaxation, bond nature alteration, charge polarization, "initial-final states" effect, etc. Secondly, according to the bond order-length-strength (BOLS) correlation[21] extended from the premise of Pauling[22] and Goldschmidt,[23] bonds between undercoordinated atoms become shorter and stronger, as confirmed in Au clusters.[24, 25] The spontaneous process of bond contraction and strengthening will cause local densification and quantum entrapment of bonding charge and binding energy and hence the positive core level shift as a consequence. Thirdly, the densely and locally entrapped bonding electrons will polarize the weakly bound $sp^2$ dangling-bond electrons. These polarized and unpaired electrons neither follow the regular dispersion relation nor occupy the allowed energy states in the valence band and below as defined by the Hamiltonian; however, they add impurity states in the vicinity of $E_F$ and follow the Dirac equation,[26] generating the STM/S probed Dirac-Fermi polarons (DFPs).[27, 28, 29, 30] Finally, the polarized and unpaired electrons with non-zero spin will in turn screen and split the crystal potential and hence generate extra component in the upper edge of the core band consequently. Figure 1a and b illustrate the BOLS correlation and the associated entrapment and polarization effect.

Analytically, the Hamiltonian for the 1s electrons of carbon can be formulated by the combination of the band theory [20] and the BOLS correlation[21] with inclusion of the polarization effect,

$$H = H_0 + H'$$
$$= \left[ -\frac{\hbar^2 \nabla^2}{2m} + V_{atom}(r) \right] + V_{cry}(r)(1 + \Delta_H)$$

**(1)**

The perturbation to the Hamiltonian $\Delta_H$ contains the following terms: [21]

$$\Delta_H = \begin{cases} \Delta_z = C_z^{-m} - 1 = E_z/E_b - 1 & (BOLS\ entrapment) \\ \Delta_p = (E_{1s}(p) - E_{1s}(0))/(E_{1s}(12) - E_{1s}(0)) - 1 & (Polarization) \end{cases}$$

$$C_z = d_z/d_b = 2/\{1 + \exp[(12-z)/(8z)]\} \quad (Bond\ contraction\ coefficient)$$

(1)

The $p$ is the coefficient of polarization to be determined from the XPS measurement. $E_{1s}(p)$ represents the peak energy of the polarization component in the XPS spectrum. $C_z$ is the coordination number (CN or z)-dependent Goldschmidt-Pauling's bond contraction coefficient, which varies only with the effective CN and has nothing to do with the dimensionality or the structure phase. $E_b$ and $d_b$ represent, respectively, the bond energy and bond length in the ideal diamond bulk. The $m$ represents the bond nature indicator.

The C 1s level of an isolated carbon atom and its shift upon the crystal potential involved follow the relations,[20, 21]

$$E_{1s}(0) = \langle 1s,i|V_{atom}(r)|1s,i\rangle$$

$$\Delta E_{1s}(z) = \langle 1s,i|V_{cry}(r)(1+\Delta_H)|1s,i\rangle\left[1 + \frac{z\langle 1s,i|V_{crym}(r)(1+\Delta_H)|1s,j\rangle}{\langle 1s,i|V_{cry}(r)(1+\Delta_H)|1s,i\rangle}\right]$$

$$\cong E_b(1+\Delta_H)\left(1 + \left(\frac{overlap\ integral}{exchange\ integral} < 3\%\right)\right) = E_z$$

(2)

where $|1s,i\rangle$ is the C 1s eigen wave function at the ith atomic site with z neighbors. $\langle 1s,i|1s,j\rangle = \delta_{ij}$ because of the strong localization of the core electrons.

For the diamond, the effective z is 12 instead of 4 because the diamond structure is an interlock of two fcc unit cells.[21] With the known C-C bond lengths in graphite (0.142 nm) and in diamond (0.154 nm), one can drive the effective z = 5.335 for an atom in the bulk graphite using the Goldschmidt-Pauling's coefficient $C_z$.

Thus, correlation between the XPS components follows the criterion,

$$\frac{E_{1s}(x)-E_{1s}(0)}{E_{1s}(12)-E_{1s}(0)} = \begin{cases} C_z^{-m} & (BOLS\ Entrapment) \\ \Delta_p + 1 & (Polarization) \end{cases}$$

(2)

The x represents z or *p*. If the polarization-entrapment coupling effect is apparent, the term $C_z^{-m}$ is then replaced by $pC_z^{-m}$, the entrapped states will be moved up from the otherwise low-z position to energy closing to the bulk component. For situations without apparent polarization, the relation evolves,

$$\frac{E_{1s}(z)-E_{1s}(0)}{E_{1s}(z')-E_{1s}(0)} = \frac{C_z^{-m}}{C_{z'}^{-m}}, \text{ or, } E_{1s}(0) = \frac{C_{z'}^m E_{1s}(z') - C_z^m E_{1s}(z)}{C_{z'}^m - C_z^m}$$

(3)

The BOLS reproduction [21] of the elastic modulus enhancement [31] and the melting point depression [32] of carbon nanotubes,[33] and the C 1s core level shifts of graphene edge, graphene, graphite, and diamond [34] has consistently confirmed that the C-C bond between two-coordinated edge atoms contracts by 30% from 0.154 nm to 0.107 nm and the bond become 252% stronger than that in diamond, giving a generalized form for the z-resolved C 1s energy shift with the optimized bond nature indicator m = 2.56, [21] as illustrated in Figure 1c,

$$E_{1s}(z) = E_{1s}(0) + [E_{1s}(12) - E_{1s}(0)]C_z^{-2.56} = 282.57 + 1.32 C_z^{-2.56}(eV)$$

**(3)**

The discovery[12] that the minimal energy (7.5 eV/bond) required for breaking a bond of the two-coordinated carbon atom near vacancy is 32% times higher than that (5.67 eV/bond) required for breaking a bond between the three-coordinated carbon atom in a suspended graphene confirms the BOLS formulation of undercoordination-induced bond strength gain.

2.2 Zone-resolved photoelectron spectroscopy (ZPS)

One can imagine what will happen in the difference between two spectra collected: (i) from the same defect-free surface at different emission angles; or (ii) from the same surface after and before the surface is being conditioned such as defect generation and chemisorption under the same measurement conditions. Upon the standard processes of background correction and spectral area normalization, the ZPS in (i) keeps the spectral features due to the skin by filtering out the bulk information as the XPS collects more information from the surface at larger emission angles.[35] Likewise, the ZPS in (ii) purifies merely the spectral features due to conditioning. The ZPS also filters out all the artifact background such as the charging effect and the "initial-final states" relaxation effect that exist throughout the course of measurements. This technique can be used for monitoring surface processes such as crystal growth, defect generation, chemical reaction, alloy formation, etc., both statically and dynamically.

III    Experimental procedures

STM/S measurements have uncovered the graphitic DFPs as extraordinarily high protrusions and resonant peak at $E_F$ from sites surrounding atomic vacancies [15, 29, 36, 37] and at the edges of monolayer graphite terraces and graphene nanoribbons (GNRs).[12, 16, 38] The DFPs demonstrate anomalies including the extremely low effective mass, extremely high group velocity, and non-zero spin, following the Dirac equation with a nearly linear dispersion crossing Fermi energy.[39, 40] Our recent density functional theory (DFT) calculations[41] revealed that the DFPs with a high-spin-density create preferably at a zigzag-GNR edge and at an atomic vacancy because of the isolation and polarization of the dangling $\sigma$-bond electrons of the identical $\sqrt{3}d$ ($d$ is the C-C bond length) distance along the edge. The locally and densely entrapped bonding electrons provide the force pinning the DFPs. However, along the armchair-GNR edge and the reconstructed-zigzag-GNR edge, the formation of quasi-triple-bond between the nearest edge atoms of $d$ distance prevents the DFPs from generation.

In order to testify the power of the ZPS and confirm the DFT calculations, we applied the ZPS to the XPS analysis in combination with the BOLS and the energy band theory to identify the nature, origin, and consequence of the graphitic DFPs with quantitative information by distilling the surface and defect states from the mixture of bulk and surface.

The XPS data were collected at different emission angles at room temperature from a graphite surface with and without artificial vacancies using the Sigma Probe Instrument (Thermal Scientific) with monochromatic Al K$_\alpha$(1486.6 eV) as the X-ray source. The XPS facility was firstly calibrated using pure gold, silver, and copper standard samples by setting the Au-4$f_{7/2}$, Ag-3$d_{5/2}$ at binding energies of 83.98 ± 0.02 eV and 368.26 ± 0.02 eV, respectively. Highly oriented pyrolytic graphite (HOPG) was cleaved using adhesive tape, and then, transferred quickly into the XPS chamber.

In order to control the generation and density of the surface atomic vacancy, we sprayed the graphite surface using $Ar^+$ ions with 0.5 keV energy incident along the surface normal. The ion dose was controlled by the sample current and the duration of spraying. The energetic $Ar^+$ bombardment creates only undercoordinated atoms at the surface without any chemical reaction in high vacuum.[42] We firstly collected the spectra at different emission angles (between the surface normal and photoelectron beam) from the defect–free surface to discriminate the skin from the bulk. We then collected the spectra from the conditioned surface at different $Ar^+$ ion doses under the same conditions and the same emission angle.

The spectra were corrected using the standard process of Shirley background correction and then the spectral peak areas were normalized using the standard process under guideline of spectral area conservation. As the total member of electrons excited from the specimen each time is proportional one to another, the spectral area was normalized to minimize the effect of scattering by the rough surface or by the surface atoms at larger emission angles. The spectra collected from the defect-free surface at larger emission angles up to 75° were then subtracted by the referential one collected from the freshly-cleaved surface at the least available emission angle (25°). For the defect density dependence, we simply repeat the process at 50° without changing the emission angle. The difference between the spectra collected at 75° from the surface with and without defect was also collected for comparison.

IV Results and discussion

Figure 2a and b show, respectively, the well-resolved XPS spectra collected (a) from the defect-free surface at different emission angles and (b) from the surface of different defect densities represented by the $Ar^+$ doses at the emission angle of 50°. The datasets gained in our results are consistent with previously reported [17, 18, 43, 44]. The angle-resolved spectra shows alight positive shift while the defect gives rise to slight negative shift of the spectra. The overall weakened intensities of the raw spectra collected at larger emission angles or from those of higher defect densities arises from the scattering loss.[45] This effect can be compensated by the spectral area normalization correction under the guideline of spectral area conservation. It is noted that the energy change rests the same for angles approaching to 75° or higher vacancy density. As the extent of the core-level shift depends on the atomic CN instead of the number of such atoms, the unapparent change of the core level energy suggests that the spectral information become dominated by the surface skins as the emission angle increases.

The ZPS in Figure 3 shows the evolution of (a) the defect-free surface and (b) the surface with different defect densities as represented by the doses of $Ar^+$ ions. The areas above and below the x-axis are, respectively, the gain and loss of the energy states under the given conditions.

According to the tight binding approximation, the separation between the spectral features and the atomic C 1s level (282.57 eV) is proportional to the bond energy. The valleys at 284.20 eV in (a) and 284.40 eV in (b) are the subtracted graphite bulk and the mixture of surface and bulk components. In addition to the spectral valleys, one entrapped peak ($T_S$) is present to the defect-free surface at the bottom edge of the C 1s band, which shifts further to a lower binding energy corresponding to the evolution of the effective atomic CN from $z \sim 4$ to $z \sim 3.2$, as the emission angle increases from 35° to 75°. As the vacancy defects are generated, the $T_S$ moves to energy even deeper and evolves into the $T_D$. The $T_D$ is associated surprisingly with the emergence of both the DFPs at the Fermi energy as identified using STM/S [12, 15, 16, 29, 36, 37, 38] and the P component at the upper edge of the C 1s band.

As the defect density is increased, the intensity of the $T_D$ component grows but remains stable in energy; the P component moves up in both energy and intensity. This finding indicates that the atomic CN has reached and stabilized at the lowest value and the extent of polarization increases with the defect density. The core level position depends on the effective atomic CN but not the density of such undercoordinated atoms. The presence and evolution of the $T_S$ to the $T_D$ component is referred to as a positive core level shift due to the surface- and defect-induced quantum entrapment making the C 1s band deeper when the atomic CN is decreased, which verify further Goldschmidt-Pauling-Feibelmans' premise of bond contraction and the BOLS anticipation. The presence of the P states is referred to as the negative core level shift, which results from the screening of the crystal potential by the presence of the DFPs at the defect sites. The DFPs screen and split partially the crystal potential and hence the core band into the $T_D$ and the P components. However, such a process of screening and splitting does not happen to the defect-free surface because of the lacking of the unpaired-dangling bond electrons at the smooth surface.

**Figure 4** and **Table 1** summarize our ZPS findings from graphite surfaces. The ZPS for the clean surface is obtained by differentiating the XPS data collected at 75° from that collected at 25°. The defect ZPS is the difference between the two spectra collected at 75° from the surface after and before high-density defect generation. The insets show the zones (in color) dominating the extra states in each case. The atomic CN for the outermost layer of graphite is estimated as 3.1 closing to the ideal case of 3.0 of graphene. The atomic CN for the vacancy extends from 2.2 to 2.4, which indicates that the next nearest neighbors to the vacancy contribute to the XPS feature. It is exciting that the atomic CN of atoms annexed the vacancy are compatible to that of the GNR edge of 2.0 because of the weak interface interaction. From **Figure 4** and the BOLS numerical expressions in (2) and (3), we can evaluate the length and strength of the bonds and the binding energy shift associated with the undercoordinated atoms, as summarized in **Table 1**. Consistency in the expected effective coordination and the specific energy between the current results and previous observations (Table 1 in [21] and Figure 1c) evidences sufficiently the accuracy of the derivatives. Strikingly, the ratio of the energy shift of the 2- and 3-coordinated atoms agrees exceedingly well with the trend of energy requirement for the 2- and 3-coordinated graphene bond breaking.[12]

As compared in Figure 5a and b, the high protrusions and the STS resonant peak of graphite surface atomic vacancy [15] are naturally the same to that of the graphene zigzag edge [16] despite the sharpness of the resonant $E_F$ peak. The peak sharpness is subject to the STM/S tip conditions and the measuring temperature. The STM/S similarity and the close values of atomic CNs between graphite atomic vacancy and graphene edge suggest that both are the same in nature. Therefore, we can focus on the graphite surface vacancy more conveniently to mimic the GNR edge. Figure 5c compares the density functional theory calculations of the local density of states of the vacancy and edge atoms. The polarized protrusions with alternative spin directions indicate that the z-edge and atomic vacancy share the same mechanism of Dirac-Fermi polaron generation as the isolation and polarization of the dangling σ-bond electrons of $\sqrt{3}d$ distance along the edge by the locally densely entrapped core and bonding electrons while quasi-triple bond formation between the nearest d separated σ-bond electrons along the arm-chaired edge or the reconstructed (5 or 7 atomic rings) z-edge prevents the DFPs formation[41].

V    Conclusion

We have demonstrated the power of the ZPS in gaining the local, atomistic zone selective, and quantitative information about the length and energy of the local bonds and the binding energy shift of electrons associated with the undercoordinated defect and surface atoms of graphite. It has been confirmed that the DFPs generate in the following processes: (i) the shorter and stronger bonds between undercoordinated atoms cause local densification and quantum entrapment of bonding charge and binding energy that produces the entrapped component in the core band; (ii) the entrapped core charge polarizes the unpaired dangling-bond electrons to produce the DFPs; and (iii) the DFPs in turn screen and split the potential and then generate the P component in the upper edge of the core band. For the clean surface, no polarization happens though the $T_S$ remains because of lacking of the dangling bond electrons. The ZPS is therefore demonstrated more revealing than using either of the STM/S or the XPS alone in purifying bond and electronic information limited to the atomistic selected zones.

Financial supports from MOE (RG15/09), Singapore, and NSF (Nos. 11172254, 11002121and 10802071) of China are gratefully acknowledged.

Table and Figure captions

Table 1 BOLS-ZPS derived length $d_z$, energy $E_z$, of the C-C bond and the electronic binding energy of the C 1s band $E_z$ of atoms annexed to the undercoordinated carbon atoms at the graphite skin with and without vacancy defects.

|  | z | $C_z$ | $d_z$(nm) | $E_z$(eV) | C 1s (eV) | P (eV) |
|---|---|---|---|---|---|---|
| Atom | 0 | - | - | - | 282.57 | |
| GNR edge | 2.00 | 0.70 | 0.107 | 1.548 | 285.89 | |
| Graphite Vacancy | 2.20 | 0.73 | 0.112 | 1.383 | 285.54 | 283.85 |
|  | 2.40 | 0.76 | 0.116 | 1.262 | 285.28 | |
| GNR interior | 3.00 | 0.81 | 0.125 | 1.039 | 284.80 | |
| Graphite Surface | 3.10 | 0.82 | 0.127 | 1.014 | 284.75 | |
| Graphite | 5.335 | 0.92 | 0.142 | 0.757 | 284.20 | |
| Diamond | 12.00 | 1.00 | 0.154 | 0.615 | 283.89 | |

figure 1 Illustration of (a) the "atomic CN-radius" correlation of Goldschmidt-Pauling-Feibelman with the scattered data of observations and the formation of the bond contraction coefficient;[46] (b) the BOLS correlation indicating that the shorter and stronger bonds between undercoordinated atoms cause local quantum entrapment and densification of the binding energy and the bonding and core electrons, which polarize in turn the nonbonding edge electrons;[47] and, (c) the positive core level shift of graphene edge (z=2), graphene (z=3), few-layer graphene, graphite and diamond with respect to the $E_{1s}(0)$,[21] in comparison with experimental results [34, 48].

Figure 2 The raw XPS spectra collected from (a) defect-free graphite surface at different emission angles and (b) the defect surface at 50° of different $Ar^+$ ion doses. One can hardly resolve difference in the binding energy but only the spectral intensity, which initiate the ZPS technique herewith.

Figure 3(a) The purified ZPS shows only the trapped ($T_S$) surface states evolving from z ~ 4 to z ~3 with emission angle increasing from 25° to 75° and (b) both the trapped ($T_D$) and polarized (P)

states coexist due to the vacancy defects. The $T_D$ is deeper than $T_S$, indicating the defect bonds are shorter and stronger than those of the surface. The surface bonds are shorter and stronger than those in the bulk.

Figure 4 Comparison of the purified XPS C 1s spectrum collected at 75° from the surface with ($9\times10^{14}$ cm$^{-2}$ dosed Ar$^+$ ion) and without defects. The valleys centered around 284.20 eV and 284.40 eV correspond, respectively, to the removed obvious graphite bulk and the mixture of surface-bulk information and the extra components are the energy states due to the surface skin, $T_S(z \sim 3.1)$, within the outermost atomic layer and sites surrounding vacancy defects, $T_D(z \sim 2.2 \sim 2.4)$. G denotes the bulk graphite ($z = 5.335$). The P component at the upper edge arises from the screening and splitting of the crystal potential by the DFPs that originate from the polarization of the entrapped ($T_D$ in the bottom of the core band) core electrons. Insets illustrate the emission angle and the atomic-diameter-thick zones (in blue color) dominating the extra core band components in each situation.

Figure 5 The STM/S profiles of (a) the graphite surface with and without atomic vacancies (permitted offprint from [15]) and of (c) the graphene zigzag edge (permitted offprint from [16]) in comparison with the density functional theory derived edge states of asymmetric-dumbbell shaped unpaired and polarized dangling bond electrons with spin up and down by the locally densified and entrapped bonding and core electrons [41].

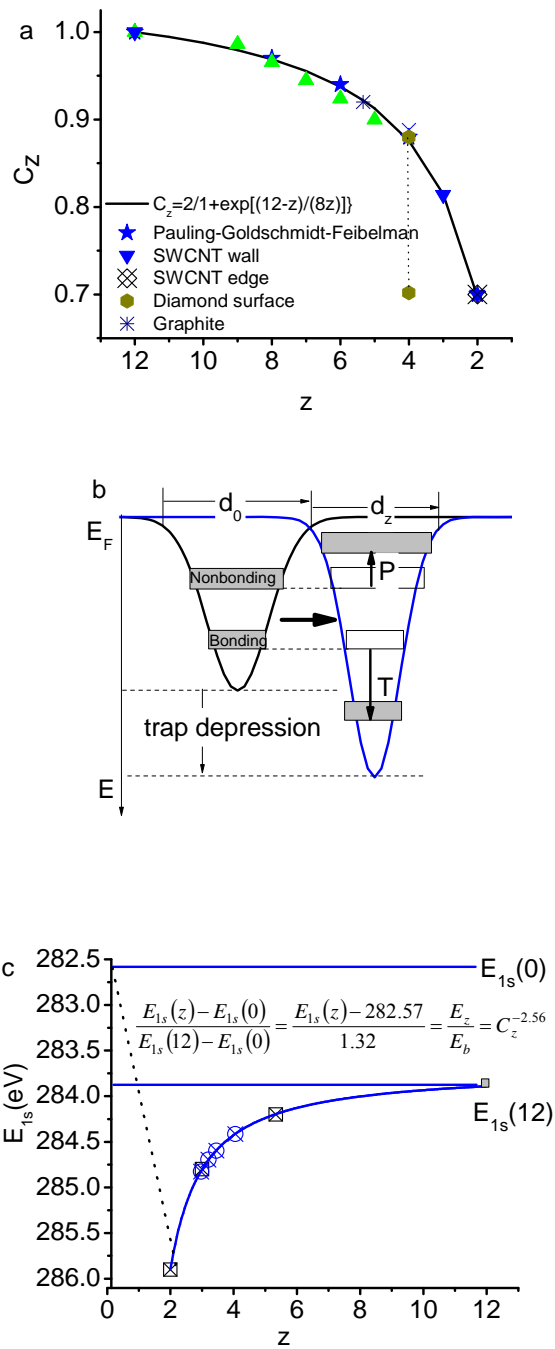

Figure 1

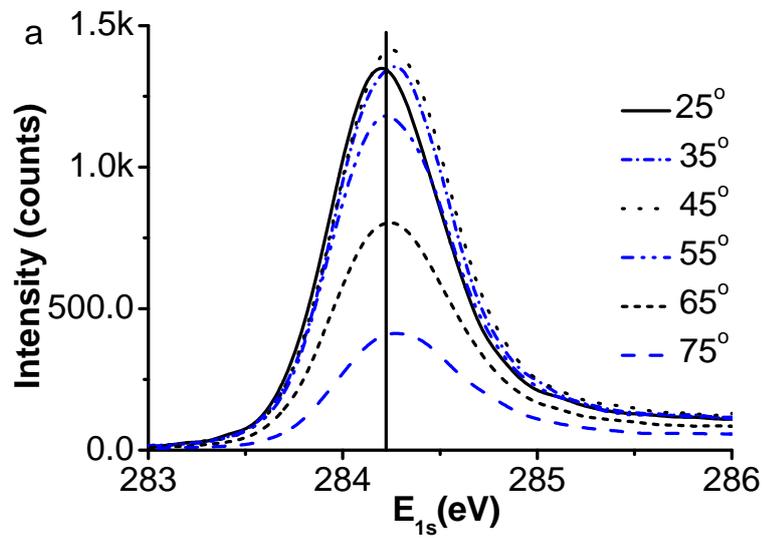
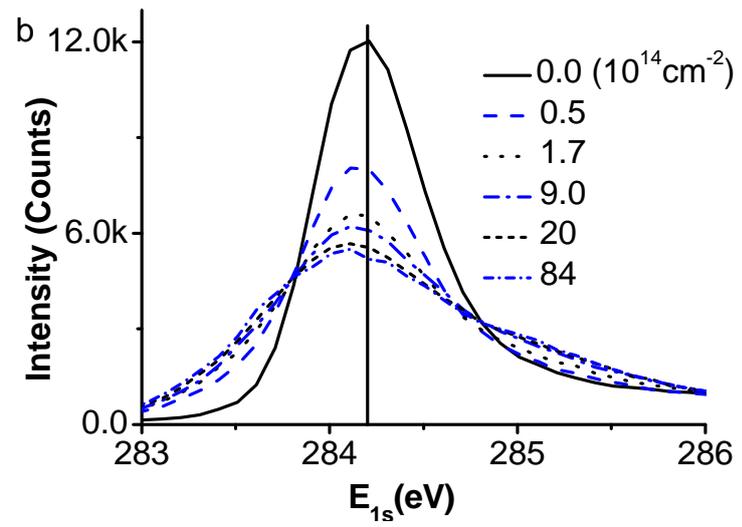

Figure 2

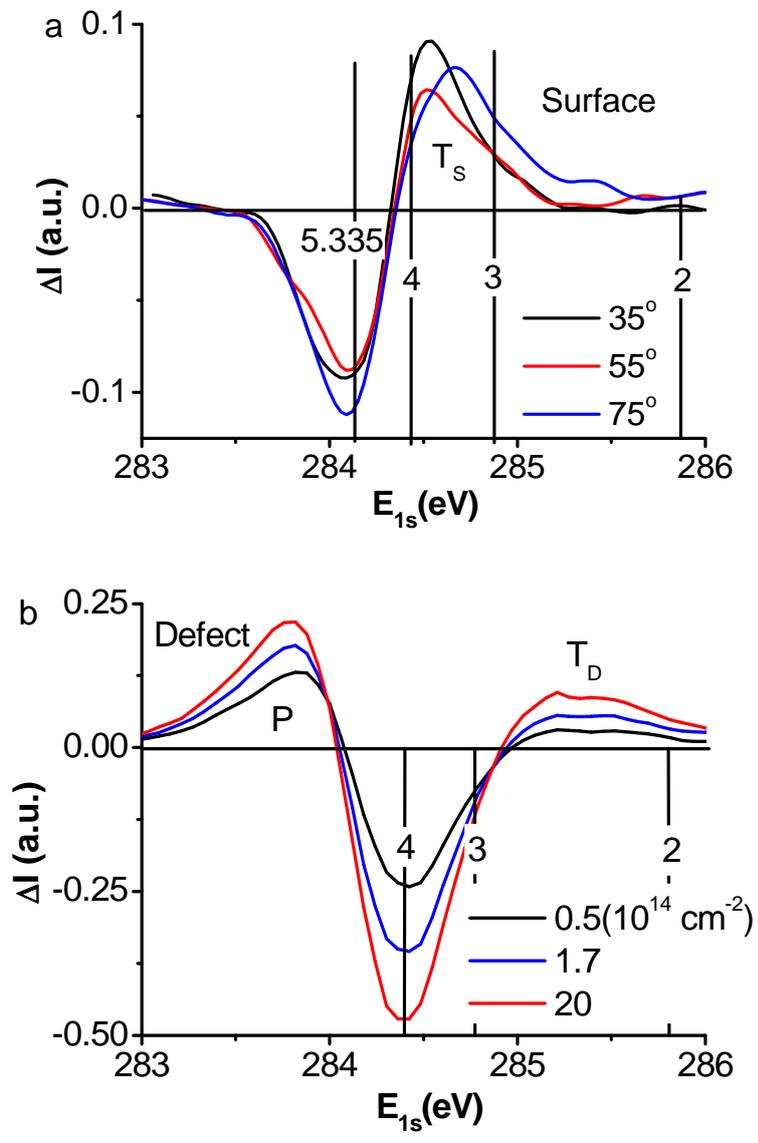

Figure 3

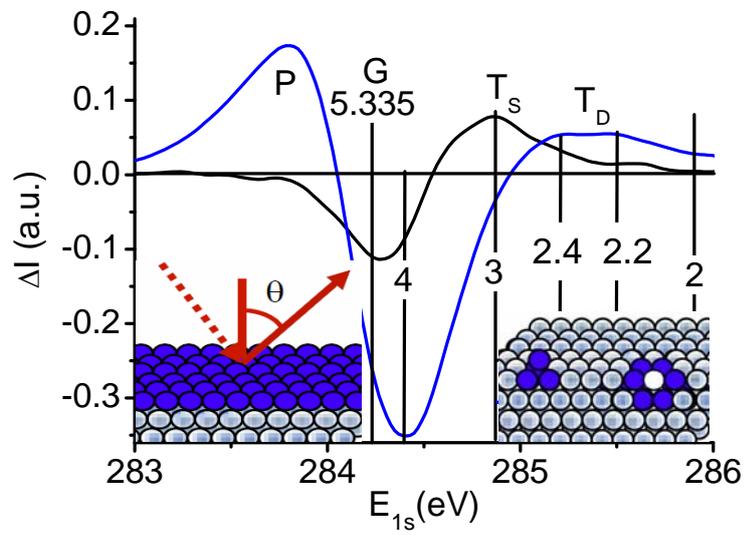

Figure 4

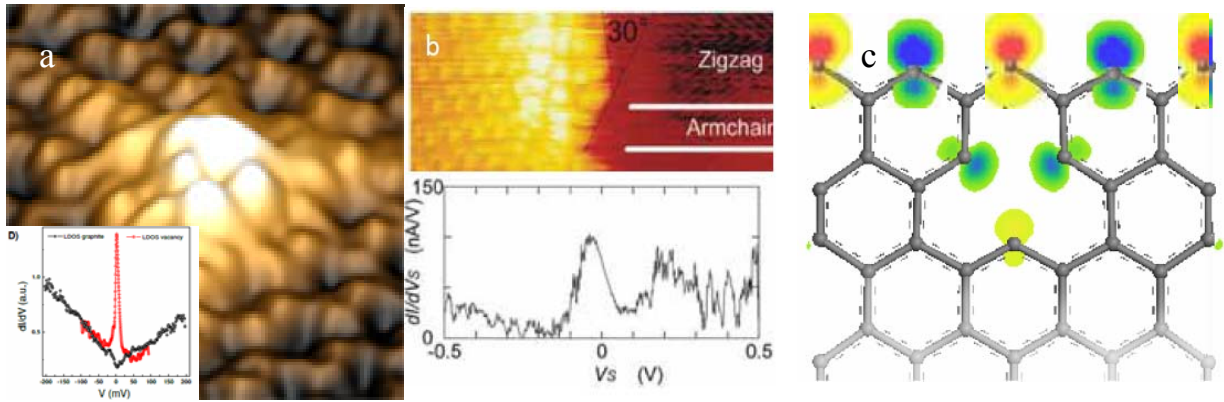

Figure 5